\documentclass{ws-procs975x65}
\usepackage{graphicx}

\def\beq{\begin{equation}}
\def\eeq{\end{equation}}

\begin{document}

\title{MOCCA survey database I. BHs in star clusters}
\author{Mirek Giersz$^*$}

\address{Nicolaus Copernicus Astronomical Center,\\
Polish Academy of Sciences, Warsaw, Poland\\
$^*$E-mail: mig@camk.edu.pl\\
}

\author{Abbas Askar} 

\address{Nicolaus Copernicus Astronomical Center,\\
Polish Academy of Sciences, Warsaw, Poland \\
and\\
Lund Observatory, Department of Astronomy, and Theoretical Physics, \\
Lund University, Lund, Sweden\\
}

\author{Jakub Klencki} 

\address{Institute of Mathematics, Astrophysics and Particle Physics,\\ Radboud University Nijmegen, Nijmegen, The Netherlands\\
}

\author{Jakub Morawski} 

\address{Astronomical Observatory, \\
Warsaw University, Warsaw, Poland\\
}

\begin{abstract}
We briefly describe and discuss the set-up of the project \textsc{Mocca Survey 
Database I}. The database contains more than 2000 Monte Carlo models of evolution of real star cluster performed with the \textsc{Mocca} code. Then, we very briefly discuss results of analysis of the database regarding the following projects: formation of intermediate mass black holes, abrupt cluster dissolution harboring black hole subsystems, retention fraction of black hole - black hole mergers, and tidal disruption events with intermediate mass black holes. 
\end{abstract}


\bodymatter


\section{Introduction}

Recent high resolution observations of globular clusters (GC) provide a very detailed picture of their physical status and show complex phenomena connected with multiple stellar populations, binary evolution, black holes (BH) and the Galactic tidal field. Despite such great observational progress there are many theoretical uncertainties connected with the origins of GCs and their primordial properties. To bridge the gap
between present-day observed star cluster properties and their properties at the time of cluster formation, we need to discriminate between different theories and models by means of numerical simulations of GC evolution. The best suited codes for such a task are N-body and Monte Carlo codes. In this paper we will describe results of Monte Carlo simulations done with the MOnte Carlo Cluster simulAtor - \textsc{Mocca}. \cite[][(and references therein)]{Hypki2013,Giersz2013} All those simulations were collected in a \textsc{Mocca Survey Database I}, which then was analyzed from the point of view of properties and evolution of different kinds of BH populations.

\begin{figure}[!ht]
\begin{center}
\includegraphics[width=5.0in]{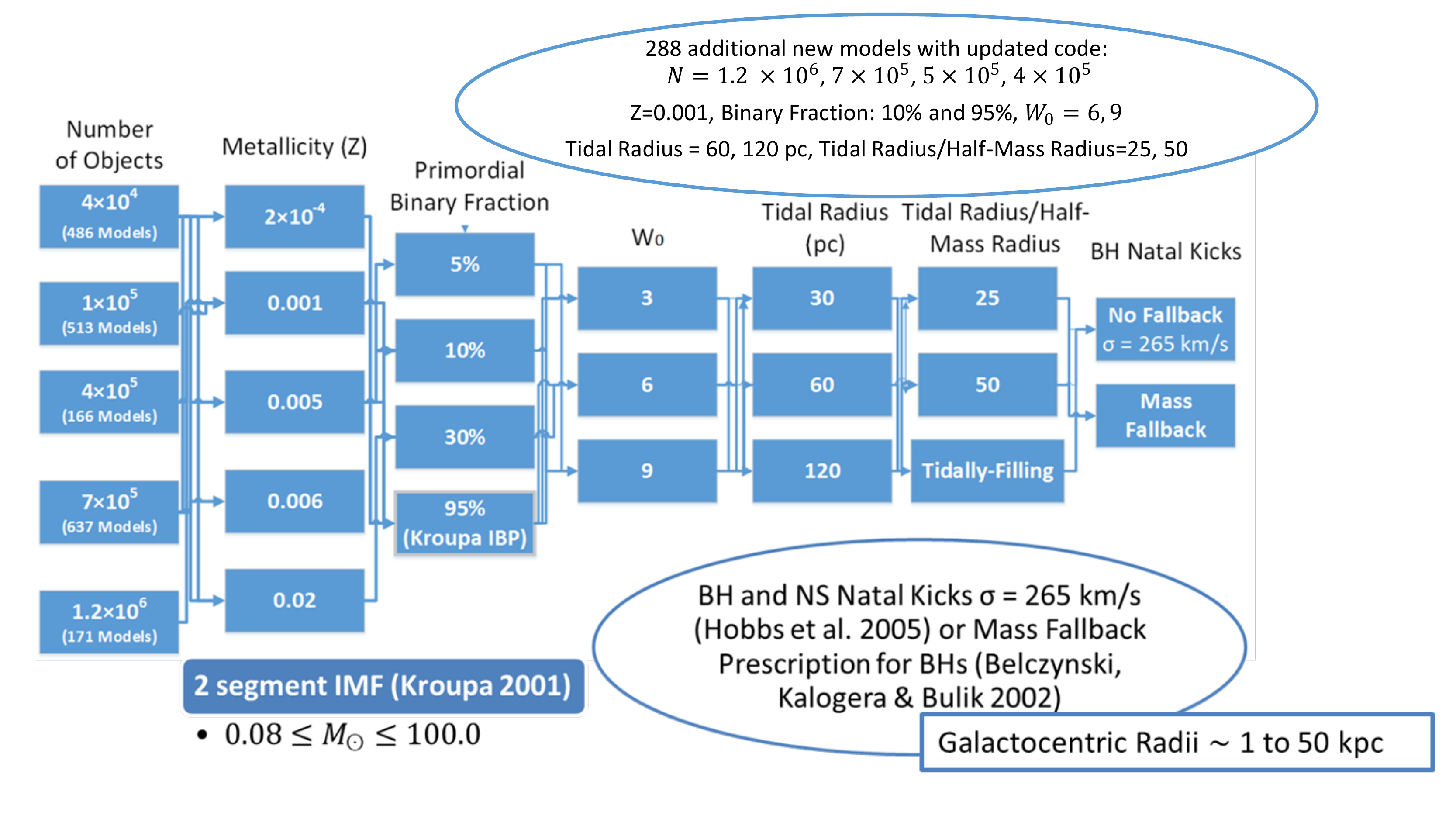}
\end{center}
\caption{The initial set-up of the \textsc{Mocca} simulations of the real star clusters stored in the \textsc{Mocca-Survey Database I}. The model parameters and relevant references are listed in the Figure. In the models the pair-instability supernovae, pulsation pair-instability supernovae, electron capture supernovae and accretion induced supernovae were not taken into account. BHs formed in the simulations have masses smaller than about $15-20~M_{\odot}$, depending on metallicity.}
\label{fig1}
\end{figure}

\section{MOCCA and MOCCA Survey Database I}

The MOCCA code is a version of the Monte Carlo codes and can be considered as so-called 'kitchen sink' code, which is able to follow most physical processes important during star cluster dynamical evolution. The \textsc{Mocca} code treats the relaxation process using the method described by Henon,\cite{Henon1971} that was significantly improved by Stod{\'o}{\l}kiewicz,\cite{Stod1982,Stod1986} and more recently by Giersz and his collaborators. \cite[(and reference therein)]{Giersz2008,Giersz2013,Hypki2013,Giersz2015} For stellar and binary evolution Jarrod Hurley's BSE code is used,\cite{Hurley2000,Hurley2002} and for the scattering experiment John Fregeau's Fewbody code.\cite{Fregeau2004} The realistic description of an escape process in a tidally limited cluster is done on the basis of the Fukushige \& Heggie theory.\cite{Fukushige2000} The \textsc{Mocca} code provides as many details as \textsc{N-body} codes. It can follow evolution and movement of particular objects. The \textsc{Mocca} code is extremely fast. It needs about a day to complete an evolution of a real size globular cluster. So, instead of just one N-body model, hundreds or thousands models can be computed with different initial conditions. The \textsc{Mocca} code is ideal either for dynamical models of a particular cluster or for large surveys. 

The \textsc{Mocca Survey Database I}\cite{Askar2017} contains about 2000 models of GCs with different initial masses, structural and orbital parameters. The brief description can be found on Fig.~\ref{fig1} and more details is given in Askar et al., Table I.\cite{Askar2017}

As it was pointed out by Askar et al.,\cite{Askar2017} it can be assumed that the \textsc{Mocca Survey Database I} cluster models are more or less representative of the Milky Way GC population.

\section{Intermediate Mass BH Formation in GCs}

In the literature, so far, there were four possible groups of scenarios proposed for intermediate mass BH (IMBH) formation in GCs:  Direct collapse of very massive Population III stars proposed by Madau and Rees, \cite{Madau2001} runaway merging of very massive MS stars in dense young star clusters, first discussed by Portegies Zwart et al.,\cite{PortegiesZwart2004} accretion of the residual gas on stellar mass BHs formed from the first generation stars recently proposed by Leigh et al.\cite{Leigh2013} and the scenario based on results of \textsc{Mocca} simulations of evolution of dense stellar systems proposed by Giersz et al.\cite{Giersz2015}. In this scenario an IMBH is formed because of buildup of BH mass solely due to mergers in dynamical interactions and mass transfers in binaries. This scenario is a much expanded and refined scenarios proposed ealier by Miller \& Hamilton \cite{Miller2002} and by Leigh et al.\cite{Leigh2013}

In the \textsc{Mocca Survey Database I} there were about 460 IMBHs formed out of 2000 models, and they formed even in very low N models, as small as consisting of 40000 objects. As it is described in detail in Ref.~\refcite[(see Fig.~6 there)]{Giersz2015} there are generally two regimes of IMBH formation: 1) very fast an IMBH mass buildup (FAST) starting from the very beginning of the cluster evolution, which requires very large initial central densities. In a time of about one Gyr, an IMBH mass grows to up a few tens of thousands $M_{\odot}$, and 2) slow IMBH mass buildup (SLOW) starting later on in the cluster evolution, usually, around the core collapse. It requires rather moderate initial central densities, and masses of IMBHs are quite moderate, from a few hundred up to a few thousand $M_{\odot}$. The process of IMBH formation is highly stochastic. The larger the initial cluster concentration, the earlier, faster and with higher probability an IMBH will form. 

Here is the detailed, slightly updated comparable to Ref.~\refcite{Giersz2015},  description of the FAST and SLOW scenarios:
\begin{enumerate}
\item SLOW and FAST formation scenarios;
\begin{enumerate}
\item SLOW scenario - either a single BH is left after the early phase of SN explosions (SNe), or a single BH is formed via mergers or collisions during dynamical interactions, usually around the core collapse time. The central density has to be greater than about $10^5~M_{\odot}/pc^3$;
\item FAST scenario - several dozen/hundreds BHs remain in the system after the early phase of SNe, and form a dense central subsystem. The central density must be extremely high, greater than $10^8~M_{\odot}/pc^3$, for an IMBH to form. Alternatively, all BHs are quickly and efficiently removed from the system via dynamical interactions. If at least one remains, then the SLOW scenario is followed;
\end{enumerate}
\item Initial mass buildup of IMBH progenitors:
\begin{enumerate}
\item If the cluster density is large enough, the collisions between main sequence (MS) stars lead to formation of very massive MS stars, hundreds of $M_{\odot}$. If such a star collides with a BH then a very massive BH (already an IMBH) is formed. 
\item If cluster density it is not large enough MS does not collide efficiently and stellar mass BH will be formed, because of stellar evolution and strong stellar winds. 
\end{enumerate}
\item BHs are the most massive objects in GC, so they quickly form a binary via a three-body interaction; 
\item Further BH binary evolution is due to dynamical interactions with other binaries and stars, or because of gravitational wave radiation (GW);
\begin{enumerate}
\item orbit tightening leading to mass transfer from MS/Red Giant/Asymptotic Giant Branch companions;
\item exchanges and collisions, leaving the binary intact;
\item total collisions during dynamical interactions or GW mergers - in this case, the binary is destroyed and only a BH is left;
\end{enumerate}
\item Newly created single BH quickly forms a new binary via another three-body interaction, which is further free to undergo subsequent dynamical interactions with other single and binary stars, and the process repeats. In this way, the BH mass steadily increases. 
\end{enumerate}

Total collisions are the most important ingredient of an IMBH formation in the SLOW scenario. In this way BH mass can steadily increase and BH binaries are not kicked out from the system due to dynamical three- or four-body interactions. That is the reason why IMBH formation is so stochastic in the SLOW scenario.

\begin{figure}[!ht]
\begin{center}
\includegraphics[width=4.0in]{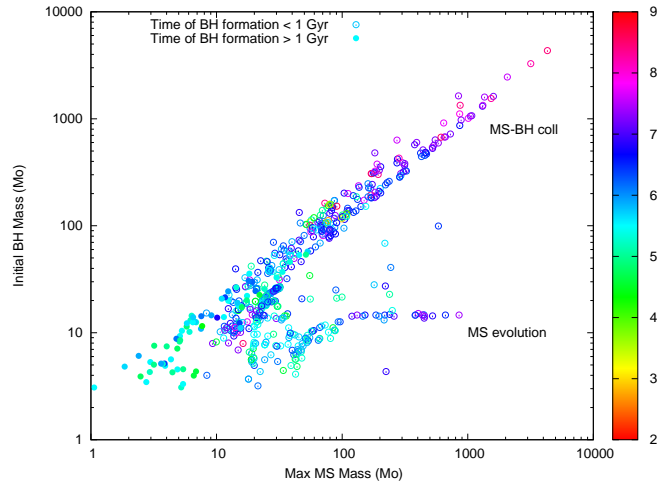}
\end{center}
\caption{Mass of just formed BH as a function of mass of MS star just before a supernova explosion. Open circles - formation of BH before 1 Gyr. Filled circles - formation of BH after 1 Gyr. The color bar on the right side shows the initial cluster central densities in $M_{\odot}/pc^3$. }
\label{fig2}
\end{figure}

As it can be seen in Fig~\ref{fig2}, there are clearly visible two channels of dynamical interactions leading to formation IMBH seeds: runaway MS star mergers and BH formation because of collisions with a BH, and runaway mergers and BH formation because of pure stellar evolution. The larger the cluster density the larger the maximum mass of runaway merged MS star and consequently the larger the BH mass (collision product of stellar mass BH and very massive MS star). Less dense models form lower mass BHs, which later in the course of evolution will substantially grow. Very massive BHs are preferentially formed for very dense models at the very beginning of the cluster evolution. We should here caution readers, the above picture relies on very uncertain properties and evolution of extremely massive MS stars formed in runaway collisions and on the amount of accreted mass onto stellar mass BH during collision with extremely massive MS star. Such collision is more similar to the common envelope phase than to direct collision of similar mass BHs and a MS stars, in which probably only small amount of mass can be accreted onto BH. We checked, that in the case of only 10\% of accreted mass an IMBH is still formed, so the FAST scenario seems to work in real physical systems.

We would like to stress that the presented scenarios for IMBH formation in GCs, in particular the SLOW scenario, do not require any specific conditions, unlike other scenarios proposed in the literature. IMBH formation occurs solely via binary dynamical interactions and mass transfer in binaries. The FAST scenario is more probable to occur in galactic nuclei, or in extremely dense star clusters.

\section{BH-BH Merger Gravitational Wave Radiation Kicks}

We used \textsc{Mocca Survey Database I} to estimate GW kick retention fraction of BH-BH merger products. We found about 4500 such mergers. The amplitude of the GW recoil kick velocity depends on the spin magnitudes, degree of misalignment between BH spins and the binary angular momentum and on the BH mass ratio. The kick velocities were calculated according to Ref.~\refcite{Baker2008} and final spins according to Ref.~\refcite{Rezzolla2008}. We checked many different assumptions about the BH spins: random, constant equal to 0.5, or a function of metallicity and initial stellar mass.\cite{Belczynski2017} There are two classes of mergers: mergers in 'primordial' binaries (binaries which keep the same stars during the whole evolution), which have only small spin misalignments, and 'dynamical' mergers in binaries which were involved in strong interactions, e.g. exchanges or disruptions, for which spin directions were distributed randomly. The evolution of each BH was tracked. Each product of BH-BH merger, also a BH, was assigned with a recoil kick velocity. Merger products with velocities greater than the cluster escape velocity were removed from any further interactions. Those merger products that remained in the cluster were assigned with a modified spin value, computed according to Ref.~\refcite{Rezzolla2008}, and could contribute to next generations of BH-BH mergers. 'Dynamical' mergers were mainly mergers with IMBHs and have small mass ratio, so nearly 0.7 of them were retained in the system. For 'primordial' binaries the retention fraction seems to saturate at about 0.2. This is connected with the fact that most 'primordial' BH-BH binaries consist of low mass BHs, which according to Belczynski et al.\cite{Belczynski2017} have large spins. On average, we should expect that about 0.3 mergers will be retained in the system. This result does not strongly depend on the assumed initial BH spin distributions.

The retention fraction is a strong function of time. In the case of 'dynamical' mergers, it is changing from the initial values of 0.5-0.6, when the IMBH mass is still relatively small, and steadily increasing up to 0.9 in the later stages of the cluster evolution, when the mass ratios in IMBH-BH mergers drop down to 0.01 - 0.001 values. The retention fraction for 'primordial' binaries substantially increases after about 1 Gyr due to decrease of the mass ratio. It seems that first binaries with mass ratio close to 1 are merged and later on  binaries with smaller and smaller mass ratios start to merge. The retention fraction for models with BH subsystems is very small and equal to about 0.1. This is because such models have relatively low concentration and low escape velocity. There is a clear correlation, the larger the cluster concentration the larger the retention fraction. Interested readers, we refer to the paper by Morawski et al.\cite{Morawski2018} where they can find all details connected with this work. 

We would like to stress that our results confirm analytic and semi-analytic results obtained earlier by many authors, but for the first time real simulation data was used to obtain the BH merger retention fraction and its dependence on the different cluster evolutionary scenarios and global cluster parameters. This approach is not fully self consistent, but it is the first step to fully integrate BH merger kicks in dynamical evolution of GCs. 

\begin{figure}[!ht]
\begin{center}
\includegraphics[width=4.0in]{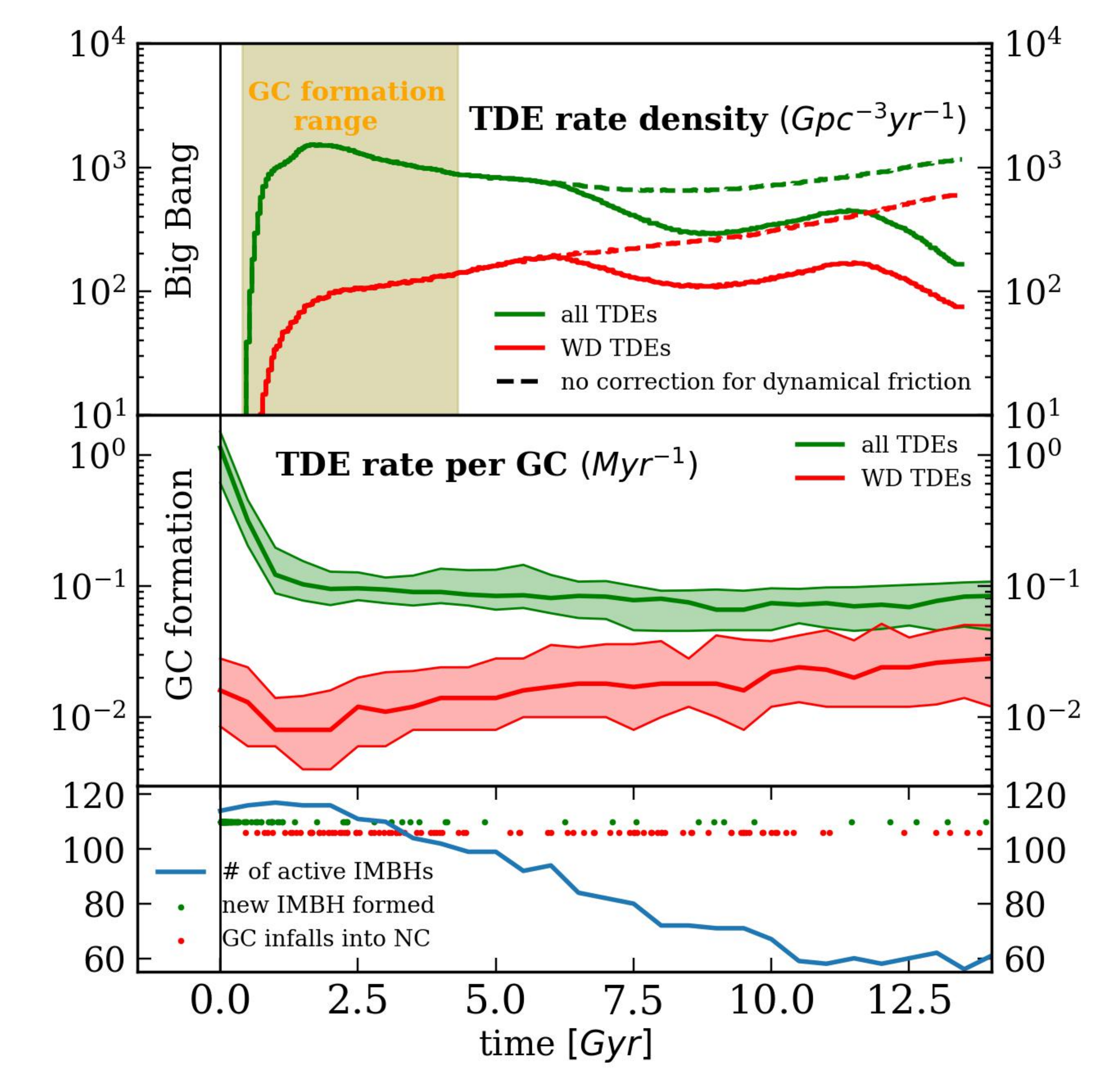}
\end{center}
\caption{The TDE rates as a function of the GC age. Red lines - rates for WDs. Green lines - rates for WDs and MS stars. Dashed line - for all models, solid line - only for models which do not migrate to the galactic center. Top panel - TDE density rate ($Gpc^{-3}yr^{-1}$). Middle panel - TDE rate per GC ($Myr^{-1}$). Bottom panel - number of active GCs. Red dots - time when GC merge with nuclear star cluster, green dots - time when new IMBH is formed.}
\label{fig3}
\end{figure}

\section{Tidal Disruption Events with IMBHs} 

Formation of IMBHs in the MOCCA models and their subsequent mass buildup have to be connected with tidal disruption of star intruders, so called tidal disruption events (TDEs). We analyzed the \textsc{Mocca Survey Database I} looking for disruptions of white dwarfs (WD), MS stars or other luminous stars. We found 344755 WD-IMBH type events, 750753 MS-IMBH type events and 42934 other-IMBH type events. The work is still in progress and we would like only to summarize briefly a very preliminary results. Most TDEs are formed in massive GCs with relatively small galactocentric distances. To form an IMBH in the FAST scenario a very large central density is needed. Such density can be achieved easily in massive GCs with small tidal radii (small galactocentric distances). Due to dynamical friction, some of such GCs will migrate towards to galactic center and merge with the nuclear cluster (NC). From that point, our models of GC evolution are no longer applicable. Interestingly, any TDE events associated with IMBHs hosted by those GCs would be observable in the galactic nuclei. In any case, in order to calculate the IMBH TDE rate in actual GCs we exclude from our computations those GC models that have fallen into the NC. In the Fig.~\ref{fig3} we show the TDE rates computed assuming that all TDEs outside the galactic center can be observed. Present day TDE rate density is smaller by factor of about 3--4 when the evolution of GCs in the galactic environment (in a Milky-Way type galaxy) is taken into account. Also the number of "active" GCs in which TDEs are happening is much smaller now than was, showing strong influence of galactic environment on the observed TDEs. Interestingly, IMBHs formed in the SLOW scenario are responsible for only a small fraction of possible TDEs. The obtained TDE rates are comparable to other theoretical estimates.

In the future work we are planning to populate galaxies in the local universe with MOCCA GCs models and estimate the local TDE rates.

\begin{figure}[!ht]
\begin{center}
\includegraphics[width=4.0in]{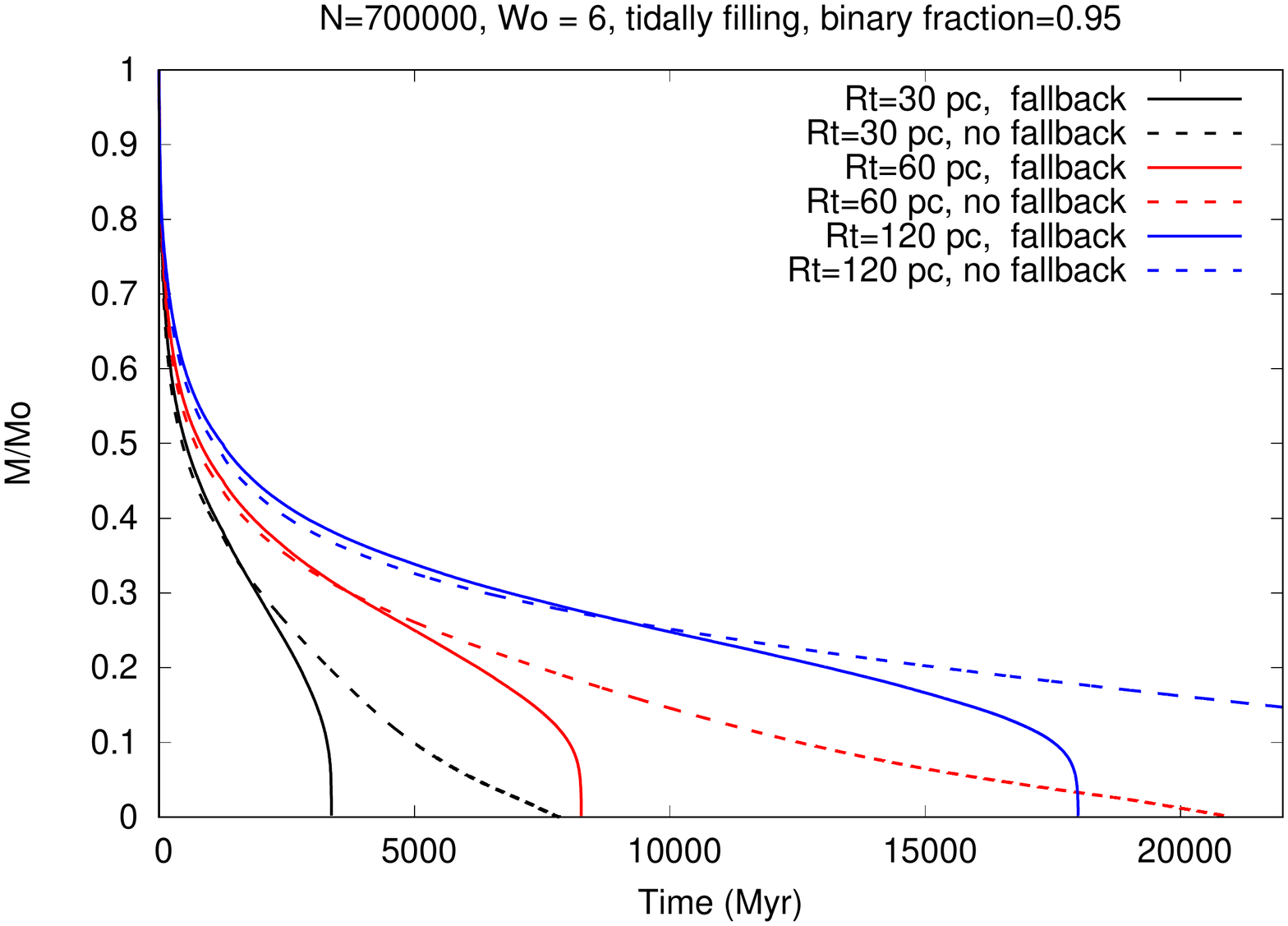}
\end{center}
\caption{Evolution of the fraction of cluster bound mass as a
function of time for tidally filling \textsc{Mocca} cluster models with
700000 objects (stars and binaries), $W_0$ = 6 and binary fraction
equal to 0.95, for different tidal radii and SNe natal kicks mass fallback
set to ON or OFF.}
\label{fig4}
\end{figure}

\section{BH Subsystem and a Third Cluster Dissolution Mechanism}

We used the \textsc{Mocca Survey Database I} to investigate the dissolution process for dynamically evolving star clusters embedded in an external tidal field, with focus on the presence and evolution of a stellar-mass BH subsystem. We argue that the presence of a BH subsystem can lead to the dissolution of tidally filling star clusters and this can be regarded as a third type of cluster dissolution mechanism, in addition to well known mechanisms connected with strong mass loss due to stellar evolution and mass loss connected with the relaxation process. As it can be seen in Fig.~\ref{fig4} for models with mass fallback ON (high BH retention fraction), the third process is characterized by abrupt cluster dissolution connected with the loss of dynamical equilibrium. The abrupt dissolution is powered by the strong energy generation from a massive stellar-mass BH subsystem accompanied by tidal stripping. We argue that such a mechanism is universal and should also work for tidally under-filling clusters with top-heavy IMF. Observationally, star clusters which undergo dissolution powered by the third mechanism would look as 'dark clusters' i.e. composed of stellar mass BH surrounded by expanding halo of luminous stars, \cite{BanerjeeKroupa2011} and they should be different from 'dark clusters' harbouring an IMBH as discussed by Ref.~\refcite{Askar2017a}. An additional observational consequence of an operation of the third dissolution mechanism should be larger than expected abundance of free floating BHs in the Galactic halo. Interested readers, we refer to the paper by Giersz et al.\cite{Giersz2019} where they can find all details connected with this work. 

\section*{Acknowledgments}

MG was partially supported by the Polish National Science Center (NCN) through the grant UMO-2016/23/B/ST9/02732. AA is currently supported by the Carl Tryggers Foundation for Scientific Research through the grant CTS 17:113 and was partially supported by NCN, Poland, through the grants UMO-2016/23/B/ST9/02732. AH was supported by Polish National Science Center grant 2016/20/S/ST9/00162. This work benefited from support by the International Space Science Institute (ISSI), Bern, Switzerland, through its International Team programme ref. no. 393 The Evolution of Rich Stellar Populations \& BH Binaries (2017-18). 


\end{document}